\documentclass[12pt]{article}
\usepackage{amsfonts}

\begin{document}

\textheight 24cm
\textwidth 16cm
\oddsidemargin 0.5cm
\evensidemargin 0.5cm
\topmargin=-1cm
\footskip 1cm
\parskip=1ex

\renewcommand\theequation{\thesection.\arabic{equation}}

\newcommand{\eqn}[1]{(\ref{#1})}
\newcommand{\be}{\begin{equation}}
\newcommand{\ee}{\end{equation}}
\newcommand{\bea}{\begin{eqnarray}}
\newcommand{\eea}{\end{eqnarray}}
\newcommand{\bean}{\begin{eqnarray*}}
\newcommand{\eean}{\end{eqnarray*}}
\newcommand{\nn}{\nonumber}

\renewcommand{\thefootnote}{\fnsymbol{footnote}}


\newcommand{\opname}[1]{\mathop{\rm #1}\nolimits} 
\newcommand{\Tr}{\opname{Tr}} 
\newcommand{\tr}{\opname{tr}} 

\def\slash#1{\rlap/#1}
\newcommand{\alg}{{\cal A}} 
\newcommand{\Aslash}{A\mkern-11.5mu/\,} 
\newcommand{\barox}{\mathrel{\overline\otimes}} 
\newcommand{\bra}[1]{\langle{#1}\vert} 
\newcommand{\stroke}{\mathbin{\vert}} 
\newcommand{\braket}[2]{\langle#1\stroke#2\rangle} 
\newcommand{\del}{\partial} 
\newcommand{\delslash}{{\partial\mkern-9mu/}} 
\newcommand{\Dslash}{D\mkern-11.5mu/\,} 
\newcommand{\eps}{\varepsilon} 
\newcommand{\Hilb}{{\cal H}} 
\newcommand{\ket}[1]{\vert{#1}\rangle} 
\newcommand{\row}[3]{{#1}_{#2},\dots,{#1}_{#3}} 
\newcommand{\thalf}{{\textstyle\frac{1}{2}}} 
\newcommand{\tihalf}{{\textstyle\frac{i}{2}}} 
\newcommand{\tquarter}{{\textstyle\frac{1}{4}}} 
\newcommand{\vev}[1]{\langle#1\rangle} 
\def\<#1,#2>{\langle#1\stroke#2\rangle} 

\newcommand{\complex}{{\mathbb C}} 
\newcommand{\quater}{{\mathbb H}} 
\newcommand{\unit}{{\mathbb I}} 
\newcommand{\natu}{{\mathbb N}} 
\newcommand{\real}{{\mathbb R}} 
\newcommand{\inte}{{\mathbb Z}} 

\def\up#1{\leavevmode \raise.16ex\hbox{#1}}
\newcommand{\npb}[3]{{\sl Nucl. Phys. }{\bf B#1} \up(#2\up) #3}
\newcommand{\plb}[3]{{\sl Phys. Lett. }{\bf #1B} \up(#2\up) #3}
\newcommand{\revmp}[3]{{\sl Rev. Mod. Phys. }{\bf #1} \up(#2\up) #3}
\newcommand{\sovj}[3]{{\sl Sov. J. Nucl. Phys. }{\bf #1} \up(#2\up) #3}
\newcommand{\jetp}[3]{{\sl Sov. Phys. JETP }{\bf #1} \up(#2\up) #3}
\newcommand{\rmp}[3]{{\sl Rev. Mod. Phys. }{\bf #1} \up(#2\up) #3}
\newcommand{\prd}[3]{{\sl Phys. Rev. }{\bf D#1} \up(#2\up) #3}
\newcommand{\ijmpa}[3]{{\sl Int. J. Mod. Phys. }{\bf A#1} \up(#2\up) #3}
\newcommand{\mpla}[3]{{\sl Mod. Phys. Lett. }{\bf A#1} \up(#2\up) #3}
\newcommand{\prl}[3]{{\sl Phys. Rev. Lett. }{\bf #1} \up(#2\up) #3}
\newcommand{\physrep}[3]{{\sl Phys. Rep. }{\bf #1} \up(#2\up) #3}
\newcommand{\jgp}[3]{{\sl J. Geom. Phys.}{\bf #1} \up(#2\up) #3}
\newcommand{\journal}[4]{{\sl #1 }{\bf #2} \up(#3\up) #4}

\hyphenation{geo-me-try}
\hyphenation{Liz-zi}


\def\cstars{$C^*$-algebras }
\def\cstar{$C^*$-algebra }
\def\iff{\Leftrightarrow}

\thispagestyle{empty}
\setcounter{page}{0}
\begin{flushright}
DSF/9-99\\
\hfill hep-th/9902187\\
February 1999
\end{flushright}
\vspace{.5cm}
\begin{center}{\Large \bf Strings, Noncommutative Geometry\\
~ \\ and \\ ~ \\
the Size of the Target Space
}
\end{center}
\vspace{1cm}
\centerline{\large Fedele Lizzi\footnote{fedele.lizzi@na.infn.it}
}
\vspace{.25cm}
\begin{center}

~\\ ~ \\
{\it Dipartimento di Scienze Fisiche, Universit\`a di Napoli
Federico II, \\ and INFN, Sezione di Napoli\\
Mostra d' Oltremare pad. 20, I-80125, Napoli, Italy.}
\end{center}
\vspace{1.5cm}
\begin{abstract}
We describe how the presence of the antisymmetric tensor (torsion) on the
world sheet action of string theory renders the size of the target space a
gauge non invariant quantity. This generalizes the $R\leftrightarrow 1/R$
symmetry in which momenta and windings are exchanged, to the whole
$O(d,d,\inte)$. The crucial point is that, with a transformation, it is
possible always to have all of the lowest eigenvalues of the Hamiltonian to
be momentum modes. We interpret this in the framework of noncommutative
geometry, in which algebras take the place of point spaces, and of the
spectral action principle for which the eigenvalues of the Dirac operator
are the fundamental objects, out of which the theory is constructed. A
quantum observer, in the presence of many low energy eigenvalues of the
Dirac operator (and hence of the Hamiltonian) will always interpreted the
target space of the string theory as effectively uncompactified.
\end{abstract}
\newpage

\section{Introduction}

It has been known for some time in string theory that the size of the
target space is not an invariant concept. A symmetry, called T-duality,
exchanges the theory of closed strings compactified in a tiny box, of the
size a small fraction of Planck length, with the theory of strings living
in large universe, of size the inverse of the tiny box (times the square of
the Planck length). This is a consequence \cite{tduality} of a symmetry of
the spectrum of the Hamiltonian of the theory, which remains invariant
under an exchange between the lattice, which defines a toroidal
compactification, with its dual lattice. This is equivalent to an exchange
between the momenta of theory (quantized in units of the inverse of the
radius), with the winding modes, closed strings which stretch across the
torus. As the latter are also quantized, but in units of the size of the
torus, target spaces with vastly different sizes are identified. While at
first sight this may seem a very curious result, the interpretation
\cite{BV} only refers to elementary concept of quantum mechanics. The
argument of Brandenberger  and Vafa is that position is just a derived
concept, as the Fourier transform of momentum spaces, and in string theory
a different choice is also possible, namely considering eigenstates of the
winding. If the compactification radius is of the order of Planck length,
the two choices are equivalent, but for a very large radius the eigenvalues
of momentum are nearly continuous, while the ones of winding are far apart,
the first one appearing  at a very large energy, it is therefore difficult
to make ``localized wave packets'' with the Fourier transform of winding.
Conversely, with a small radius of compactification, it is the winding
which gives the possibility to create localized wave packets.

The exchange between the compactification lattice and its dual is however
just a part of a larger group of symmetries, namely $O(d,d,\inte)$, and
some of these transformation mix momentum and winding in a non trivial way.
The purpose of this paper to argue that a quantum observer will always
"see" an uncompactified space, provided the Hamiltonian has a spectrum
containing many small eigenvalues (we will make this more precise later).
We will accomplish this using the tools of noncommutative geometry, that is
we will consider string theory as a noncommutative geometry. We take the
example of the (flat) toroidal compactification. Our aim is to just show a
mechanism, and this must be understood as an example, the principles behind
this mechanism are much more general. The large scale topology of the
universe is of course unknown. We will be only concerned with physical
interpretation, and the mathematics used will only have an ancillary
purpose. The non mathematically inclined reader should bear with the
mathematical parts, the main concepts should understandable even without
the formal apparatus. The mathematically inclined reader should be tolerant
of the (mis)use of mathematics in this paper.

The crucial observation \cite{lizziszaboprl,lizziszabocmp} is that, in the
noncommutative geometry of strings, the group $O(d,d,\inte)$ is a part of
the group of gauge transformations. Also important is the fact that, with a
proper transformation, it is possible to have all of the lower eigenvalues
to be in the momentum sector. We will use these observations in the context
of Noncommutative geometry, for which the important feature is the spectrum
of the (generalized) Dirac operator. The action in noncommutative geometry
is a {\em spectral action}, it depends essentially from the lower (smaller
than a cutoff parameter) eigenvalues of the Dirac operator. The
noncommutative geometry of strings is given by the vertex algebra of the
vertex operators of the theory. The Dirac operator can be used to identify
the low energy (tachyonic sector) of the theory. Using these ingredients,
and the fact that the topology and metric of a space in noncommutative
geometry are described solely in terms of operators on an Hilbert space, we
will argue that a quantum observer will see the space as effectively
uncompactified.

The plan of the remainder of the paper is as follows. In section 2 we
introduce the string theory of interest, and the $O(d,d,\inte)$ symmetry of
interest. In section 3 we will describe the configuration space experienced
by a quantum observer in the framework of noncommutative geometry. The
relevant concept of noncommutative geometry are briefly introduced. In
section 4 we introduce the noncommutative geometry of the string theory of
interest, and in particular the Dirac operator, showing also how duality
symmetries are gauge transformation leaving the spectrum of the Dirac
operator invariant. In section 5 we introduce briefly the spectral action
principle and argue how a low energy quantum observer will necessarily see
an uncompactified space. Section 6 contains conclusions and open problems.

\section{Target Space Symmetry of a String Theory}

Consider the bosonic (sector of the) string, with the target space
compactified on a $d$ dimensional torus. That is, consider a linear
$\sigma$ model compactified on $\real^d$ quotiented by an abelian infinite
group (a lattice) $\Gamma$ generated by $d$ generators $e_i$. Then the
space ${\cal T}_d\equiv\real^d/\Gamma$ is a $d$ dimensional torus. On
$\Gamma$ we define an inner product of the generators, which provides a
metric (of Euclidean signature) on ${\cal T}_d$:
\be
\vev{e_i,e_j}\equiv g_{ij} \ \ . \label{metric}
\ee
The lengths of the vectors $e_i$ (here and in the following we work with
units such that Planck's length equals unity) give the classical `size' of
the target space, which is compact and periodic in all dimensions. We will
see that the presence of torsion and quantum mechanical considerations will
considerably alter this classical picture.

The dual lattice $\tilde\Gamma$ is spanned by the basis $e^i$
with (we implicitly complexify $\Gamma$ and extend the product):
\be
\vev{e^i,e_j}=\delta^i_j \ \ .
\ee
The inner products of the $e^i$'s define a metric which is the inverse of
$g_{ij}$, that is
\be
\vev{e^i,e^j}\equiv g^{ij} \ \ .
\ee
Notice that, if all of the $e_i$ are quantities of order $R$, with $\det g$
of order $R^d$, then the `size' of the dual lattice is (very roughly
speaking) of order $1/R$. In this sense, if to a given lattice corresponds
a large universe, to its dual it will correspond a small one, the dual
torus $\tilde{\cal T}_d$.
 This
conclusion is valid however, even in this rough form, {\em only in the
absence of torsion} in the action \cite{DouglasHull}.

Classically the  string is described by a two dimensional nonlinear $\sigma$
model, whose fundamental objects are the Fubini--Veneziano fields, which,
for the case of a closed string\footnote{In this paper we will consider
only the simplest case of closed strings. The presence of open strings,
D-branes etc. will make this structure probably much richer, but
will not be considered in this paper.}, are
\be
X^i(\tau,\sigma)=x^i+g^{ij}p_i\tau+g^{ij}w_i\sigma
+\sum_{k\neq0}\frac1{ik}~\alpha^{(\pm)\mu}_k~e^{ik(\tau\pm\sigma)} \ \ ,
\label{FubVen}
\ee
where $x$ represents the centre of mass of the string, $p$ its momentum and
$w$ is the winding number, that is the number of times the string wraps
around the direction defined by the $e_i$. Notice that because the space is
compact, the momentum is quantized, and in fact it must be
$p\in\tilde\Gamma$, while the winding number must belong to the dual
lattice $w\in\Gamma$. If the size of the target space is extremely large,
then the momentum will have a spectrum with very close eigenvalues, a
nearly continuous spectrum, while the windings will have values far apart.
But apart from these scale considerations, the role of $p$ and $w$ in
\eqn{FubVen} is symmetric. In the following we will concentrate on the zero
modes of the string, mostly ignoring the oscillator modes. These are
internal excitations of the other string, and are not sensible to the
target space in which the strings live, and will therefore in general play
no role in this paper. Moreover, the oscillators describe excitations which
are starting at the Planck mass, while most of our considerations relate to
the low energy sector of the theory.

The action of the model is that of a two dimensional nonlinear $\sigma$
model\footnote{Other terms, such as a dilatons term, are possible, but we
will not consider them here.}:
\be
S=\frac 1{4\pi} \int d\sigma d\tau \left(\sqrt{\eta}
\eta^{\alpha\beta}\del_\alpha X^i g_{ij} \del_\beta x^j +
\varepsilon^{\alpha\beta} b_{ij}\del_\alpha X^i\del_\beta X^j\right) \ \ ,
\ee
where $\eta$ is the world sheet two dimensional metric, $G$ is the matric
defined in \eqn{metric}, and $b$ is an antisymmetric tensor which represent
the `torsion' of the string.

We can perform a chiral decompositions of the $X$'s defining:
\be
X^i_\pm(\tau\pm\sigma)=x^i_\pm+g^{ij}p_j^\pm(\tau\pm\sigma)
+\sum_{k\neq0}\frac1{ik}~\alpha^{(\pm)i}_k~e^{ik(\tau\pm\sigma)} \ \ .
\label{chiralmultfields} \label{FubVenpm}
\ee
The zero modes $x^i_\pm$ (the centre of mass coordinates of the string) and
the (centre of mass) momenta $p_i^\pm=2\pi p_i\pm(g-\mp b)_{ij} w^j$ are
canonically conjugate variables,
\be
[x^i_\pm,p_i^\pm]=-i\delta_i^j \ \ ,
\label{cancommmomx}
\ee
with all other commutators vanishing. The left-right momenta are
\be
p^\pm_i=\mbox{$\frac1{\sqrt2}$}\left(p_i\pm\langle e_i,w\rangle
\right)
\label{momlattice}
\ee
The $p^\pm$'s belong to the lattice:
\be
\Lambda=\tilde\Gamma\oplus\Gamma
\ee
We can therefore define the fields $X=X_++X_-$, and we may equally well
define $\tilde X\equiv X_+-X_-$, whose zero mode we will indicate as
$\tilde x$.

The Hamiltonian of the theory (limiting ourselves to the zero mode part)
is very symmetric in this chiral decomposition:
\bea
H&=&\frac 1{2} \left( (2\pi)^2 p_ig^{ij}p_j + w^i(g-bg^{-1}b)_{ij}w^j +4\pi
w^ib_{ik}g^{kj}p_j\right) \label{spectrum}\\ &=&\frac 12 (p_+^2+p_-^2)
\eea
Since the momenta and the windings belong to a lattice, the spectrum is
discrete.

The symmetry which exchanges the lattice with its dual is called T-duality
\cite{tduality,GPR}. It corresponds to an exchange of the momentum quantum
number with the winding. The zero mode corresponding to $x$, the position
of the centre of mass of the string, is exchanged with its dual $\tilde x$.
As heuristically discussed in the introduction (and as we will again argue
below), two target spaces related by a T-dual transformation are
indistinguishable at low energies. In the torsionless case $b=0$ this
corresponds to an exchange of $g$ with its inverse $g^{-1}$, and the change
of size of the target space in which the radius $R\to 1/R$. In the presence
of torsion the exchange is $g^{-1}\leftrightarrow g-bg^{-1}b$ and
$bg^-1\leftrightarrow
-g^{-1}b$, and it depends crucially on the values of the $b_{ij}$. In the
toroidal case it is possible to exchange only some of the generators of the
lattice with their duals, giving rise to a group of factorized T-dualities.

Even factorized T-duality is a subgroup of a larger group of symmetries
which leaves the spectrum invariant, the full group is in fact
$O(d,d,\inte)$ \cite{Narain,GPR}. It is generated from three kinds of
transformations:
\begin{itemize}

\item[-] The factorized dualities we have already discussed.

\item[-] The changes of base of the lattices, made via a matrix which
belongs to $G(d,\inte)$, the group of integer valued matrices of unit
determinant.

\item[-] The transformation $b_{ij}\to b_{ij}+c_{ij}$ with $c$ an
antisymmetric tensor with integer entries.

\end{itemize}
There is a further $\inte_2$ symmetry obtained exchanging $\sigma$ and
$\tau$ on the world sheet, but this last symmetry does not affect the
target space.

Let us analyse in some details the third transformation. It changes the
components of the antisymmetric second rank tensor $b_{ij}$ by the addition
of an arbitrary integer constant. This transformation does not change the
lattice $\Gamma$, as it operates only on the antisymmetric tensor $b$. It
does however change the momenta conjugated to the zero modes of $X$ and
$\tilde X$. In particular, in the spectrum \eqn{spectrum}, the relative
contribution of the momenta conjugated to $x$  (represented by the first
term,) with respect to the windings, conjugated to $\tilde x$, and the
mixed term will change. Since this is a symmetry of the spectrum, the set
of numbers which are the eigenvalues is of course unchanged, but the
distribution in the three terms changes. By choosing the integers which
compose the antisymmetric tensor $b$ arbitrarily large, we can make the
contribution of the second and third term arbitrarily large. in other words
we concentrate the lowest eigenvalues of the Hamiltonian in the momentum
part. In other words the low energy spectrum is made only of the momentum
eigenvalues. The lattice is still the same, but the strings are extremely
twisted, and we have transferred the lowest eigenvalues of the energy from
winding to momentum. This relatively simple observation is the key of our
construction. In the following sections we will argue that in this case a
quantum observer will observe a spacetime in which the actual radii of
compactification will effectively be unobservable. Roughly speaking, a low
energy strings for which in the original (small radius) lattice had a
combination of momentum and winding, will now be twisted in such a way that
it will appear to have just momentum, it is like the lattice ``repeats
itself over and over''.

Again, as in the case of the of the $R\leftrightarrow 1/R$ symmetry, we
have to ask ourselves `what is position'? `How do I measure it'? And using
the same heuristic arguments of \cite{BV}, we can think of making wave
packets using superpositions of the eigenvalues of the momentum, in the
case of large torsion the eigenvalues of momentum are continuous for all
practical purposes, therefore the superposition will have the character of
a uncompactified space, rather than a string moving on a lattice. And this
will be the situation until energies in which the new eigenvalues (coming
from windings or the oscillatory modes) start to play a role. In the
following section we will argue this from a quantum mechanical point of
view, using the tools of noncommutative geometry.

\section{Configuration Space in Quantum Mechanics.}

In this section we will discuss the role of the classical configuration
space in quantum mechanics. This is an extremely complicated subject, which
would require a full understanding of the foundations of quantum mechanics,
and its classical limit. Since this full understanding is lacking, we only
point out some facts which we feel relevant. We will use the language and
formalism of noncommutative geometry, but will keep the formalism to a
minimum, hoping to make the relevant physical principles emerge as clearly
as possible. A discussion along these line appeared in \cite{Balbaby}, a
rigorous treatment of quantum mechanics based on an algebraic approach can
be found, for example, in Haag's book \cite{Haag}, while the main reference
for noncommutative geometry is the book by Connes \cite{book}, other useful
reviews are \cite{Landi,Madore,ticos}.

Consider, in the following, a purely quantum observer, that is somebody
making experiments with a set of operators which form an algebra. For
example bounded operators constructed from $p$ and $x$. Despite its
historical name not necessarily all self adjoint operators on an Hilbert
space can be considered to be related to an experimental procedure. In fact
the programme of {\em rigged} Hilbert space, (for reviews see
\cite{rigged}) is based on a particular choice of a subspace of the Hilbert
space, on which some operators act in such a way to have only states with a
finite energy. Here we will take a related but different point of view,
based more on the algebra of operators.

The usual way to construct quantum mechanics for the motion on a manifold
$M$, is to consider the Hilbert space $L^2(M)$, and an algebra of (bounded)
operators acting on it. Of course the Hilbert space does not carry any
information on $M$ (all separable Hilbert spaces are isomorphic), but the
information on the topology of $M$ can be recovered by considering the
algebra of {\em position operator}, that is, the algebra of
continuous\footnote{In the following we will consider $M$ compact,
therefore continuous functions are bounded as well.}, complex valued,
functions on $M$, seen as operators on $L^2(M)$, with a norm given by the
maximum of the modulus of the function.

The set of continuous complex valued functions on a topological space form
in fact an {\em abelian} \cstar, and according to a series of theorems due
to Gel'fand and Naimark (for a review see \cite{FellDoran}), it is possible
to recover the original space in an unique way from the knowledge of just
the abstract algebra. The points of the topological space in this case are
the set of irreducible representations of the algebra, and the topology of
the space can be recovered as well. Alternatively, the space can be
reconstructed as the set of maximal ideals. Ideals are subalgebras such
that the product of one of their elements by any element (of the whole
algebra) still belongs to the ideal. Maximal ideals are ideals which are
not contained in any other ideal (except the trivial ideal of the whole
algebra). The relevant example is the set of functions vanishing at a
point.

There is a third way to identify the topological space corresponding to a
given \cstar algebra, it is via the {\em pure states} of the algebra. A
state is a  map from the algebra into complex numbers with the properties
of being positive definite, of unit norm and such that:
\be
\Psi:\alg\to\complex~~~~;~~~~\Psi (a^*a)\geq 0\ ,\forall a\in
\alg,~~~||\Psi||=1\ .
\ee
It is another result of Gel'fand that all \cstar can be represented as
bounded operator on an Hilbert space $\cal H$, then vectors
$|\psi\rangle\in{\cal H}$ define states via the expectation values. States
are however much more general. We denote the space of states by ${\cal
S}(\alg)$. Since
\be
\lambda \Psi + (1-\lambda) \Phi \in {\cal S}(\alg)\ ,~~~~\forall  \Psi, \Phi
\in
{\cal S}(\alg),\ \lambda\in[0,1]
\ee
the set of all states of an algebra $\alg$ is a convex space. Being a
convex space ${\cal S}(\alg)$ has a boundary whose elements are called {\it
pure states}. The `delta-functions', seen as maps from the algebra of
continuous function into complex numbers:
\be
\delta_x(f)\equiv f(x)~~~~,~~~~ f\in\alg \label{deltafun}
\ee
are examples of pure states. Namely, a state is called pure if it cannot be
written as the convex combination of (two) other states. We can therefore
reconstruct the space as the space of pure states, which coincides with the
set of irreducible representations, and hence the space of characters.
Moreover, in the commutative case, it coincides with the space of maximal
ideals of $\alg$. The reconstruction of the underlying topological space
from pure states is actually quite close in spirit to quantum physics.

Although the Hilbert space ${\cal H}$  has been introduced to represent our
algebras, we will give it a physical meaning and see it as the space of
wave functions required by quantum mechanics. To reconstruct a topological
space, all we need is then an abelian subalgebra of the algebra of
observables. At this, purely topological level, there are however many
ambiguities (to some extent similar to the polarization choices in
geometric quantization), one could choose the algebra of momentum
operators, or combinations of position and momentum etc.

We will consider the configuration space of a quantum mechanical space
therefore not as a set of points (and relations such as topology or a
differential structure), but rather as an abelian $C^*$-algebra. This is
our starting point. The Hilbert space could also be easily constructed a
posteriori by giving a sesquilinear form (a scalar product) on the algebra,
and completing it under the norm given by this product. Other choice for
the Hilbert space are possible, a relevant one for instance is the space of
spinors. A quantum observer will have at his disposal, among the bounded
operators on the Hilbert space, an abelian subalgebra  that he will
identify with the continuous function on his space.

The ``size'' of this configuration space is given by a (generalized) Dirac
operator $D$, a self adjoint, densely defined, compact resolvent operator
on the Hilbert space\footnote{We are assuming at this point that the
Hilbert space is a space of spinors.}. The distance between two points $x$
and $y$ is given, in terms of $D$, by Connes' formula \cite{book}:
\be
d(x,y)=\sup_{||[D,a]||\leq 1}\left|a(x)-a(y)\right| \ \ . \label{distances}
\ee

Since we are interested in doing physics with noncommutative geometric
tools, we should be able to introduce also potential and covariant
derivatives as operators on ${\cal H}$. Also in this respect the Dirac
operator plays a crucial role. With it is in fact possible to represent
differential forms as bounded operators \cite{book}.

Given an abstract algebra of $p$-forms:
\bea
\omega&=&\sum_i a_{0_i} da_{1_i} \ldots da_{p_i} \nn\\
d\omega&=&\sum_i da_{0_i} da_{1_i} \ldots da_{p_i}
\eea
we define a linear representation of $\omega$ and $d\omega$ as bounded
operators:
\bea
\pi(\omega) &=&\sum_i a_{0_i} [D,a_{1_i}] \ldots [D,a_{p_i}] \nn\\
\pi(d\omega)&=&\sum_i [D,a_{0_i}] [D,a_{1_i}] \ldots [D,a_{p_i}] \ \ .
\label{forms}
\eea
Since it may occur that $\sum_i {a_0}_i d{a_1}_i \ldots d{a_p}_i=0$ while
$\sum_i d{a_0}_i d{a_1}_i \ldots d{a_p}_i\neq 0$, care has to taken in
quotienting out these forms (the so called ``junk'' forms), for details see
for example \cite{book,Landi}. For instance, in the usual commutative case,
in which the algebra is the one of complex valued functions on an a
manifold, the Hilbert space is the one of spinors on spacetime and the
Dirac operator is the usual one $D=\gamma^\mu\del_\mu$, the forms $dx^\mu$
are represented by $\pi(d x^\mu)=\gamma^\mu$. In going to higher forms one
has to retain only the antisymmetric part of the product of gamma
matrices. In the following we will omit to explicitly indicate the symbol
$\pi$ when we talk of forms, which we will assume always represented on
${\cal H}$. Once we have defined forms we can then define connections
(generic hermitean one forms):
\be
A=\sum_ia_i[D,b_i]
\ee
and a covariant Dirac operator
\be
D_A=D+A \label{covder}
\ee
The curvature also can be defined.
\be
F_A=[D,A]+A^2 \label{curv}
\ee
Connections and curvature transforms properly under a gauge group. In fact
in noncommutative geometry gauge transformation have a nice
characterization as the unitary transformation of the algebra into itself
(or the inner automorphisms). If in fact we conjugate all of the element of
the algebra by an unitary element ${\cal A}\to U^{-1}{\cal A}U$, the
physics must remain unchanged. This means that the differential forms
defined in \eqn{forms} have to transform properly. In fact for a matrix
algebras this invariance is exactly the invariance for a unitary gauge
group.

A final ingredient is the integral, since we are constructing a formalism
based on algebra of operators, without the geometrical concepts of points
etc, the best characterization of integration is as a {\em trace}. In
general for commutative algebras is possible to show that the integral of a
function is the (properly regularized) trace of this function (represented
as an operator in the Hilbert space) times $|D|^{-d}$. We define the
Dixmier trace $\tr_\omega L$ of an operator $L$, with eigenvalues
$\lambda_n$, with $\lambda_{n+1}\geq\lambda_n$, to be:
\be
\tr_\omega L=\lim_{N\to\infty}{1\over \log N}\sum_{n=1}^N \lambda_n
\ee
For the algebra of continuous functions on a $d$-dimensional compact
manifold $M$, this definition yields \cite{book}
\be
\int_M f(x)~d^px =\tr_\omega~ f |D|^{-d} \ \ .
\label{intdirac}\ee

We have therefore equipped our quantum observer with a series of tools
suited to him: algebras of  operators, traces etc. In the commutative case
these tools reconstruct the usual differential geometry, but we have
defined them in such a way that hey can be used without any reference to an
underlying ``commutative'' geometry. If we are in a commutative case, the
quantum observer has therefore at his disposal an algebra of observables,
in this algebra he recognizes an abelian subalgebra, that he calls the
space on which he lives, and with formula \eqn{distances} he calculates
distances, metric etc. The set of an Hilbert space ${\cal H}$, a
\cstar realized as operators on ${\cal H}$ and a Dirac operator is called
the {\em Spectral Triple}.

There is the possibility that the quantum observer finds himself on a {\em
noncommutative space}. That is, among his set of quantum observable he
does not identify an abelian algebra giving him the configuration space,
he can however define (in an approximate sense perhaps) some sort of
``noncommutative'' space, the algebra corresponding to it is however non
abelian, usually a deformation, governed by a small parameter, of an
abelian algebra. This is, for example, the situation envisaged in
\cite{DFR}, in which the algebra of position operators is noncommutative.

There is also an intermediate possibility, suppose for example that the
Hilbert space the observer has at his disposal is that of spinors with an
index, which transform under the fundamental representation of $SU(n)$. In
this case the algebra of ``position" operators is actually made of
functions from the manifold to $n\times n$ matrices. This is obviously a
noncommutative algebra, so the Gel'fand--Naimark theorem (at least in the
commutative form we have enunciated) does not apply, it is nevertheless
obvious that the configuration space is the manifold $M$ all the same. The
choice of abelian subalgebras (diagonal matrices) would create various
identical copies of the manifold. The question is easily resolved noticing
that the algebra of $n\times n$ matrices has only one nontrivial
irreducible representation, or rather, all representations are unitarily
equivalent. Therefore the set of irreducible representations (up to unitary
transformations) of the algebra is still in a one to one correspondence
with the underlying manifold $M$. The noncommutativity of the algebra
however makes it impossible the identification of points with pure states,
there is in fact a $n$ dimensional sphere of pure states at each point of
space (corresponding to the various unitarily equivalent representations)
and they can be seen as an ``internal" space, points connected by a gauge
transformation.

We see therefore that there are noncommutative algebras which give (at
least at the topological level) the same geometry, this concept is
captured by the concept of {\em Morita equivalence} \cite{Rimor}.
Two Morita
equivalent \cstars have equivalent representation theories, therefore at
the topological level they will describe the same set of `points', with the
same topology.
 The
``physics'' they would describe (if we interpret them as the algebra on
space) differ therefore only in the ``internal space''.  Quantum observers
using Morita equivalent algebras of operators will therefore conclude they
are describing different theories {\em on the same manifold}.
In the fifth section we  give a criterion based on the spectrum of the
Dirac operator, with which he can identify an algebra which will give the
quantum observer his space, commutative or noncommutative.

\section{The Noncommutative Geometry of Strings}

In this section we will describe the noncommutative geometry corresponding
to the string theory of section 2. As we have seen in the preceding section
we need three ingredients for this purpose, let us construct them in turn.
We will be necessarily brief and details (and further references) can be
found in \cite{FG,FGR,lizziszaboprl,lizziszabocmp,lizziszabochaos,LLS}. We
start the construction from the Hilbert space ${\cal H}$ on which the
$X^\pm$ act as quantum operators:
\be
{\cal H}=L^2\left({\cal T}_d\times
\tilde{\cal T}_d\,,\,\mbox{$\prod_{i=1}^d\frac{dx^i\,d\tilde
x_i}{(2\pi)^2}$}\right) \otimes{\cal F}^+\otimes{\cal F}^-
\label{hilbertdef}
\ee
The ${\cal F}^\pm$ are Fock spaces on which the creation and annihilation
oscillatory modes act (and therefore will not be relevant at low energies),
while the $L^2$ space of spinors in \eqn{hilbertdef} is generated by the
canonical pairs of position and momenta zero modes \eqn{cancommmomx}, and
can be expressed in several isomorphic ways:
\bea
L^2\left({\cal T}_d\times
\tilde{\cal T}_d\,,\,
\mbox{$\prod_{i=1}^d\frac{dx^i\,d\tilde x_i}{(2\pi)^2}$}\right)
&\cong&
L^2\left(T_d\,,\,\mbox{$\prod_{i=1}^d\frac{dx^i}{2\pi}$}\right)\otimes_\complex
L^2\left(T_d^*\,,\,\mbox{$\prod_{i=1}^d\frac{d\tilde x_i}{2\pi}$}\right)\nn\\
&\cong&
\bigoplus_{w\in
L}L^2\left({\cal T}_d\,,\,\mbox{$\prod_{i=1}^d\frac{dx^i}{2\pi}$}\right)\nn\\
&\cong&\bigoplus_{p\in L^*}L^2\left(\tilde{T}_d\,,\,
\mbox{$\prod_{i=1}^d\frac{d\tilde x_i}{2\pi}$}\right)
\label{hilbertisos}
\eea
That is, we can consider the zero modes part of ${\cal H}$ as copies of the
spinors of the torus (a copy for each winding), or as copies of spinors on
the dual torus, or as tensor product of two spinor spaces, reflecting the
various choices of the spectrum.

We wish to describe {\em interacting} strings, and therefore we must use
operators which describe strings splitting, joining etc. Such operators are
called {\em vertex operators}, they form an algebra and have a
distinguished position in mathematics as well as in physics (for reviews
see for example \cite{VOA}), although  we will need very little of the
beautiful mathematical formalism of these algebras. The most important
feature is the {\em operator state correspondence}. This means that, to
each state on the string Hilbert space, it is possible to associate an
operator acting on the vacuum:
\be
V_\psi\ket{0}=\ket{\psi}
\ee
The fundamental vertex operator is the so called ``tachyonic'' vertex
operator (this can actually be a misnomer since, according to the string
theory one is considering, tachyons can be absent):
\bea
V_{q^+q^-}(z_+,z_-)&\equiv&
V(e^{-iq_\mu^+x_+^\mu-iq_\mu^-x_-^\mu}\otimes\unit;z_+,z_-)\nn\\
&=&(-1)^{q_\mu
w^\mu}:~e^{-iq_\mu^+X_+^\mu(z_+)-iq_\mu^-X_-^\mu(z_-)}~:
\label{tachyon}
\eea
where $(q^+,q^-),(r^+,r^-)\in\Lambda$, $z_\pm=^{-i(\tau\pm\sigma)}$, and
$:\cdot:$ denotes normal ordering.

Thus we take as second ingredient the vertex operator algebra. A word of
caution, in general a vertex operator algebra is not a \cstar. In fact,
operators defined as in \eqn{tachyon} are not bounded. To overcome this one
can smear them \cite{FG}, however also smeared operators are not
necessarily bounded \cite{FGR,ConstScharf}. Alternatively one can consider
a cutoff on the oscillators (effectively an ultraviolet cutoff on the world
sheet) \cite{LLS}. The algebra, seen as bounded operators on ${\cal H}$,
will then have to be completed to give a \cstar.

The last ingredient we need is a Dirac operator. In this case we can in
fact naturally define two of them:
\be
D^\pm(\tau\pm\sigma)=\sum_{k=-\infty}^\infty D
_k^\pm~e^{ik(\tau\pm\sigma)}
\label{diracrampm}\ee
where
\be
D_k^\pm=\sqrt2~\gamma_\mu^\pm\otimes\alpha_k^{\pm\mu}
\label{dirramk}\ee
with $\alpha_0^\pm=p^\pm$ and the $\gamma^\pm$ are the appropriate gamma
matrices in the $\pm$ sectors. In analogy with the definition of momentum
and winding from the $p^\pm$ we can define two operators:
\bea
D&=&D^++D^- \label{dirac}\\ \tilde D&=&D^+-D^-\label{diractilde}
\eea
The algebra of vertex operators ${\cal A}$, the Hilbert space
\eqn{hilbertdef} and $D$ are the three elements which form the
Fr\"ohlich-Gaw\c edzki spectral triple, which describes the noncommutative
geometry of interacting strings. The algebra of vertex operators is of
course non abelian and very complicated, and therefore the geometry of the
stringy spacetime is highly nontrivial. This is not surprising, we know the
extreme richness and beauty which lies behind string theory, and the
intricacies (and beauty) of vertex operator algebras are its algebraic
counterpart.

We note that, if we construct ``vertex operators'' only with $x$, we have
multiplicative operators of the kind $e^{ikx}$, which are the Fourier basis
of the algebra of position operators, which can be therefore seen as a
subalgebra of the vertex operator algebra\footnote{We stress again that
here by vertex operator algebra we need the algebra opportunely regularized
and completed.}. The spaces of the zero modes $x$ or $\tilde x$ can be
recovered using the Dirac operators as follows \cite{lizziszabocmp}. We
first define the Hilbert space obtained from the Hilbert space
\eqn{hilbertdef} simply eliminating the oscillators Fock spaces:
\be
{\cal H}_0 = {\cal P}{\cal H} \label{H0}
\ee
We then define the subalgebra ${\cal A}_0$ to be the commutant of the Dirac
operator $\tilde D$, restricted to the Hilbert space ${\cal H}_0$,
\be
{\cal A}_0={\cal P}_0\,({\rm comm}~D)\,{\cal
P}_0\equiv\{V\in{\cal A}~|~[D,V]\,{\cal P}_0=0\}
\label{A0comm}\ee
It is the largest subalgebra of $\cal A$ with the property
\be
{\cal A}_0{\cal H}_0={\cal H}_0
\label{A0def}\ee

The essence of T-duality form the noncommutative geometry point of view is
that we could have substituted the role of $D$ in \eqn{A0comm} with $\tilde
D$ as first suggested in \cite{FG}. In fact , together with $\{{\cal
A},{\cal H},D\}$, there is a totally equivalent dual spectral triple
$\{{\cal A},{\cal H},\tilde D\}$. That is, one can define an algebra
$\tilde {\cal A}$ dual of ${\cal A}$, which commutes with $D$. This explain
the symmetry. Another crucial aspect \cite{lizziszaboprl} is that this
transformation is a {\em gauge transformation} of the theory.

As we said the gauge group of a spectral triple is given by the unitary
elements of the algebra. The unitary element of a vertex operator algebra
form a very complicated group, which is the gauge group of the string.
Under a the gauge transformation the Dirac operator transforms as $D\to
U^{-1}DU$.
The point is that there are several unitary elements of the algebra
\cite{lizziszabocmp} such that
\be
\tilde D = D\to U^{-1}DU \label{diracequiv}
\ee
This shows that $D$ and $\tilde D$ {\em have the same spectrum}.
The gauge transformation \eqn{diracequiv} only relabels the eigenvalues of
the operators, calling momentum eigenvalues the ones which were winding
eigenvalues and viceversa for example.

To conclude we note that, although the construction described here refers
to the bosonic string, generalizations to other strings based on the
toroidal compactification, such as the heterotic string, are possible
\cite{szabosong}.

\section{Spectral Action and Low Energy Theories}

We want to discuss some ``low energy'' limit in string theory, and
therefore we have to specify better what we mean. In this section we will
briefly introduce the Chamseddine--Connes spectral action principle
\cite{ChamsConnes}, and will argue that low energy means considering a
theory in which only the low part of the spectrum of $D$ is considered.
This is possible because in the framework of Noncommutative geometry one
constructs a {\em spectral geometry}, in which the information is stored in
the spectrum of $D$. And low energy refers to an action in which only the
lower part of the spectrum is excited.

The spectral action principle is based on the covariant Dirac operator, and
on the variation of its eigenvalues. The action must be read in a Wilson
renormalization scheme sense, and it depends on an ultraviolet cutoff
$m_0$:
\be
S_{m_0}=\Tr\chi\left({D_A^2\over m_0^2}\right) \label{specac}
\ee
where $D_A$ is the covariant derivative defined in \eqn{covder} and
$\chi(x)$ is a function which is 1 for $x\leq 1$ and then goes rapidly to
zero (some smoothened characteristic function). The action \eqn{specac}
effectively counts the eigenvalues of the covariant Dirac operator up to
the cutoff. Considering, in fact, the eigenvalues of $D_A$ as sequences of
numbers, and these sequences as dynamical variables of euclidean gravity,
the spectral action is then the action of ``general relativity'' in this
space \cite{LandiRovelli}. The trace in the action can be calculated using
known heath kernel techniques \cite{ChamsConnes,Marseille}, and the
resulting theory contains a cosmological constant, the Einstein--Hilbert
and Yang--Mills actions, plus some terms quadratic in the Riemann tensor.
Chamseddine \cite{Chamseddine} has used the Dirac operators \eqn{dirac} in
the spectral action principle and shown that they lead to the low energy
effective string action.

Here we are not to be concerned with the details of \eqn{specac}, nor with
its results for the description of gravity and the standard model. What we
wish to stress is that such an action comes from a {\em spectral
principle}, that is, the starting point is the spectrum of an operator, and
its variations as the backgrounds fields (the one--form $A$ in this case)
change. One can ask, in fact, what is the role of the algebra in the
spectral action, as the latter depends just on the trace of the Dirac
operator. Of course the role of the algebra is in the fact that in
\eqn{specac} appears the {\em covariant} Dirac operator. And the form
$A=\sum a_i[D,b_i]$ depends on the algebra chosen. Let us now apply these
considerations to the Fr\"ohlich-Gaw\c edzki spectral triple.

The spectrum of $D$ and $\tilde D$, or of any operator obtained from them
with an $O(d,d,\inte)$ unitary transformations are the same. Let us call
$D$ for convenience the one for which the lower eigenvalues are the one
relative to momentum. Here by lower we mean the ones which are lower than
the energy of the oscillatory modes (of the order of the Planck mass
$m_p$). If the cutoff $m_0$ is lower than $m_p$, the cutoff function $\chi$
causes the projection of the operator on the Hilbert space ${\cal H_0}$
defined in
\eqn{H0}. Elements of the algebra which commutes with $D$ (such as the
elements of $\tilde{\cal A}$) will not contribute to the variations of the
action, and will therefore be unobservable. This algebra can be constructed
as the commutant of the T-dual operator $\tilde D$. This means that the
winding modes degrees of freedom are unobservable. Since the Dirac operator
has a near continuous spectrum, the tachyonic, low energy, algebra is
spanned by operators of the kind
\be
V_p=e^{ipq} \ \ ,
\ee
can be considered the Fourier modes describing
an uncompactified space.

In fact, in the spirit of section 3 of this paper, a quantum observer with
a spectral action, will have, as potentials at his disposal, only the
elements of the algebra which give low energy perturbations of the lowest
eigenvalues of $D$, always with the assumption of the cutoff  $m_0<m_p$ so
that oscillatory modes do not play a role. This is the abelian algebra of
functions on some space time. If, as we have seen, there are many low
eigenvalues, the observer will experience an effectively decompactified
space time. The algebra which he will measure will be composed of the
operators which will create low energy perturbation to $D$. At this point
we have to make the sole assumption that $D$ has a spectrum with several
small eigenvalues. In this way the quantum observer will experience a
(nearly) continuous spectrum of the momentum, the sign of an uncompactified
space.

The strings could still be seen as compactified on a ``small'' lattice, but
the presence of a very large torsion term $b$ has drastically changed the
operator content of the theory, and this has rendered space effectively
uncompactified.

\section{Conclusions}

The usual geometric notions of points, distances etc. are basically
classical. They will need to be redefined a quantum observer, dealing with
states on an operator algebra. The claim of this paper is that the torsion
term makes the observable radius of the target space a non invariant
concept, since, with a gauge transformation, it is possible to render it
arbitrarily large (under the fairly mild assumption that there are many low
energy eigenvalues of the Hamiltonian). A quantum observer will measure in
fact the Fourier transform of a near continuous spectrum, that is a
uncompactified space. We have argued this with the simplest possible
example of bosonic theory. The framework in which we investigated this has
been noncommutative geometry, and in particular the spectral action
principle.

The presence of open strings, and in particular of D-branes will probably
enrich this picture even more. In fact a theory of D-branes has a matrix
action in the large mass limit, and in turn this theory can be compactified
on a noncommutative torus \cite{CDS}, a genuine noncommutative space
\cite{Rieffel} generated by $d$ unitary generators with the relation
\be
U_iU_j=e^{i\omega_{ij}}U_jU_i \ \ .
\ee
The noncommutative Torus is also instrumental in the construction of the
T-dual action in \cite{LLS}. Different theories of branes related by
$O(d,d,\inte)$ transformation give rise to Morita equivalent noncommutative
tori \cite{morita,schwarz}.

Needless to say, there are several unanswered questions. For example, in
the formalism, there is no reason at present for which there are four
uncompactified dimensions. Or rather, in the spirit of this paper, why the
eigenvalues of the Dirac operator are such that only for four dimension the
spectrum is nearly continuous. The formalism we have used is euclidean,
therefore question such has the expansion of the universe cannot even
asked. Moreover we passed dangerously close to several fundamental aspects
of quantum mechanics. We attempted on purpose to stay away from issues such
as the semiclassical limit and the quantum theory of measurement. Those
issues are of course of paramount importance, but we decided not to
concentrate on them, giving priority to more concrete aspects of string
theory. It is our strong feeling that the tools noncommutative geometry are
the right ones to enable the description of a stringy geometry. In this
paper we have seen just an example, others still wait to be investigated.

\bigskip

\noindent
{\bf Acknowledgements.} \\ The material leading to section 3 is the result
of various discussions with A.P.~Balachandran, G.~Landi, G.~Marmo,
A.~Simoni, P.~Teotonio-Sobrinho, and various other colleagues. I would also
like to thank R.J.~Szabo for many discussions on the noncommutative
geometry of strings and G.~Landi for suggestions on a preliminary version
of the manuscript.
%
%
\bibliographystyle{unsrt}

\end{document}